\documentclass[aps,reprint,superscriptaddress,twocolumn,floatfix,showpacs,longbibliography]{revtex4-1}

\usepackage{graphicx,color}
\usepackage[utf8]{inputenc}
\usepackage{amsmath,amssymb,bm,amsfonts,dsfont,mathrsfs,amsthm}
\usepackage{braket}
\usepackage{txfonts,comment}
\usepackage[colorlinks=true,linkcolor=blue,citecolor=blue,urlcolor=blue]{hyperref}
\usepackage{soul}
\usepackage{enumitem}
\usepackage{orcidlink}
\usepackage{multirow}

\newcommand{\nn}{\nonumber \\}

\AtBeginDocument{%
    \newwrite\bibnotes
    \def\bibnotesext{Notes.bib}
    \immediate\openout\bibnotes=\jobname\bibnotesext
    \immediate\write\bibnotes{@CONTROL{REVTEX41Control}}
    \immediate\write\bibnotes{@CONTROL{%
    apsrev41Control,author="08",editor="1",pages="1",title="0",year="1"}}
     \if@filesw
     \immediate\write\@auxout{\string\citation{apsrev41Control}}%
    \fi
}%

\begin{document}

\title{Noisy Stark probes as quantum-enhanced sensors}

\author{Saubhik Sarkar\,\orcidlink{0000-0002-2933-2792}}
\email{saubhik.sarkar@uestc.edu.cn}
\affiliation{Institute of Fundamental and Frontier Sciences, University of Electronic Science and Technology of China, Chengdu 611731, China}
\affiliation{Key Laboratory of Quantum Physics and Photonic Quantum Information, Ministry of Education, University of Electronic Science and Technology of China, Chengdu 611731, China}

\author{Abolfazl Bayat\,\orcidlink{0000-0003-3852-4558}}
\email{abolfazl.bayat@uestc.edu.cn}
\affiliation{Institute of Fundamental and Frontier Sciences, University of Electronic Science and Technology of China, Chengdu 611731, China}
\affiliation{Key Laboratory of Quantum Physics and Photonic Quantum Information, Ministry of Education, University of Electronic Science and Technology of China, Chengdu 611731, China}

\begin{abstract}

Wannier-Stark localization has been proven to be a resource for quantum-enhanced sensitivity for precise estimation of a gradient field.
An extremely promising feature of such probes is their ability to showcase such enhanced scaling even dynamically with system size, on top of the quadratic scaling in time.
In this paper, we address the issue of decoherence that occurs during time evolution and characterize how that affects the sensing performance.
We determine the parameter domains in which the enhancement is sustained under dephasing dynamics.
In addition, we consider an effective non-Hermitian description of the open quantum system dynamics for describing the effect of decoherence on the sensing performance of the probe. 
By investigating the static and dynamic properties of the non-Hermitian Hamiltonians, we show that quantum-enhanced sensitivity can indeed be sustained over certain range of decoherence strength for Wannier-Stark probes. 
This is demonstrated with two examples of non-Hermitian systems with non-reciprocal couplings.
 
\end{abstract}

\maketitle

\section{Introduction}

Quantum sensing refers to the utilization of the quantum features to achieve precision in estimation problems that can go beyond the capability of classical sensors~\cite{degen2017quantum, braun2018quantum, ye2024essay}.
The precision can be quantified by the Fisher information of the sensing probe. 
The resource efficiency of the probe is quantified by the scaling of Fisher information with respect to the system size $L$, namely $L^{\beta}$. 
Here, the exponent $\beta{=}1$ represents the standard limit which is accessible with classical sensors. 
On the other hand, $\beta{>}1$ represents quantum-enhanced sensitivity, which is obtainable only by exploiting quantum features such as entanglement.  
Quantum-enhanced sensitivity was first demonstrated in interferometric setups by exploiting Greenberger–Horne–Zeilinger (GHZ) type entangled states~\cite{giovannetti2004quantum, giovannetti2006quantum, boixo2007generalized, giovannetti2011advances}, which results in $\beta {=} 2$ (Heisenberg limit).
The GHZ-based sensors are limited to phase imprinting unitary operations, difficult to scale up, and very prone to decoherence and particle loss~\cite{demkowicz2012elusive, de2013quantum}.
Quantum many-body systems provide an alternative approach for sensing~\cite{montenegro2024review}, providing more flexibility in state preparation, better scalability, and more robustness against imperfections.
Quantum criticality, as a key feature of many-body systems, has been identified as a resource for sensing. 
This includes first-order~\cite{raghunandan2018high, mirkhalaf2018supersensitive, yang2019engineering, heugel2019quantum, sarkar2024first}, second-order ~\cite{zanardi2006ground, zanardi2007mixed, zanardi2008quantum, invernizzi2008optimal, gu2010fidelity, gammelmark2011phase, skotiniotis2015quantum, rams2018limits, chu2021dynamic, liu2021experimental, montenegro2021global}, Stark~\cite{he2023stark, yousefjani2025nonlinearity, yousefjani2023long} and quasi-periodic~\cite{sahoo2024localization} localization, Floquet~\cite{mishra2021driving, mishra2022integrable}, time crystal~\cite{montenegro2023quantum, iemini2023floquet, yousefjani2025discrete, gribben2025boundary, shukla2025prethermal}, and topological~\cite{sarkar2022free, mukhopadhyay2024modular} phase transitions. 
Energy gap closing in these systems plays a key role for achieving the quantum-enhanced sensitivity. 

Criticality is typically associated with the ground state with vanishing energy gap which is generally hard to prepare due to long preparation time or thermal noise.
Moreover, criticality based sensors are also local in nature as the quantum advantage is available only in the vicinity of the transition~\cite{montenegro2021global,mukhopadhyay2024modular, mukhopadhyay2024saturable}.
Therefore, it is often desirable to create the probe state using non-equilibrium dynamics. 
This provides several advantages as initialization can be easy, enhancement can occur over a large region, and quantum resources needed for enhanced sensitivity, such as entanglement, can be generated during the dynamics~\cite{montenegro2022sequential, yang2023extractable, bhattacharyya2024restoring, balatsky2025quantum}.
In such scenarios, the evolution time is also a resource and the Fisher information scales with both time and system size as $t^{\alpha} L^{\beta}$. 
Quantum enhancement demands both the exponents $\alpha$ and $\beta$ to be greater than 1~\cite{ilias2022criticality}. 
Fundamental limits for achieving quantum enhanced sensitivity have also been identified in various forms of single-parameter quantum sensing~\cite{boixo2007generalized, abiuso2025fundamental, puig2024dynamical}.

A suitable system that displays such enhancement both in terms of equilibrium states and non-equilibrium dynamics is the Stark system on an one-dimensional (1D) tight-binding lattice with a gradient field that induces localization.
In this case, all the eigenstates show quantum-enhanced sensitivity for weak fields~\cite{he2023stark}.
The non-equilibrium dynamics of such systems has also been used for achieving quantum-enhanced sensitivity~\cite{manshouri2024quantum}.
This property makes the Stark system an excellent candidate for observing quantum-enhanced sensing on a physical platform.
Apart from photonic~\cite{peschel1998optical, jiang2022direct}, semiconductor~\cite{leo1992observation}, and superconducting qubit~\cite{guo2021observation} devices, cold atoms in optical lattices have proved to be exceptionally effective for realizing and measuring these quantum systems due to microscopic control, long coherence time, and remarkable detection schemes~\cite{dahan1996bloch, anderson1998macroscopic, meinert2014interaction, preiss2015strongly}. 
Nevertheless, an unavoidable source of decoherence arises from the spontaneous emission events due to coupling to the vacuum  modes of the electromagnetic field. 
Therefore, it is necessary to study the noisy dynamics to characterize the sensing capability correctly.
Indeed, finding a quantum sensor with reasonable robustness against decoherence is highly desirable. 
Every sensing scheme shows different degrees of resilience against decoherence.  
For instance, GHZ-based sensors are extremely susceptible to decoherence and  lose their quantum advantage quickly in the presence of noise~\cite{demkowicz2012elusive, de2013quantum}. 
Criticality-based sensors are expected to be more resilient against noise~\cite{montenegro2024review}. 
However, the range over which the quantum advantage can be achieved is model dependent and has to be specified for each sensing scheme. 
Given the effectiveness of the Stark systems, we try to quantify their performance in the presence of decoherence.

The open system dynamics is intricately connected to an effective description of the evolution under non-Hermitian (NH) Hamiltonians~\cite{ashida2020non, bergholtz2021exceptional, okuma2023non}. 
NH systems have also been explored for sensing, although mostly for boundary perturbations~\cite{wiersig2014enhancing, chen2017exceptional, hodaei2017enhanced, liu2016metrology, yu2020experimental, mcdonald2020exponentially, budich2020non, koch2022quantum}.
However, it is also possible to estimate global Hamiltonian parameters with Heisenberg precision~\cite{sarkar2024critical}.
With recent developments in the study of NH Stark systems~\cite{longhi2009bloch, longhi2014exceptional, longhi2015bloch, zhang2024dynamics, zhang2024extended}, it is important to recognize the connection between the open system Stark probe and the corresponding NH sensors.

In this paper, we study the effect of decoherence on the dynamics and quantify how it affects the sensitivity of the Stark probe.
Such open system dynamics is related to an effective NH description and therefore, we consider a similar evolution under a NH Hamiltonian to investigate the quantum sensing advantage.
The NH cases are further studied for the eigenstates that show quantum-enhanced sensitivity. 
The paper is structured as follows.
In Sec.~\ref{Sensing}, we give an overview of a general single parameter estimation problem.
In Sec.~\ref{Model}, we introduce the Stark system and point out the sensing features both in the static and dynamic cases.
In Sec.~\ref{Dephase}, we quantify the effect of dephasing in the dynamical sensing capability of a Stark probe.
Section~\ref{Traj} explains the connection between the open system dynamics and NH Hamiltonians with quantum trajectory method.
In Sec.~\ref{NH}, we look at NH Hamiltonians as an exact description of open system dynamics and analyze the sensing performance both in terms of the eigenstates and dynamics.
This is exemplified with lattice systems with non-reciprocal and unidirectional hopping in Sec.~\ref{HN} and Sec.~\ref{Lattice}, respectively.
After showcasing the precision achieved for these case in Sec.~\ref{SNR}, we conclude in Sec.~\ref{Conclusion}.

\section{Parameter estimation} 
\label{Sensing}

We start with an overview of the single parameter estimation that is considered in this paper.
Here, the unknown parameter $\lambda$ is encoded onto a quantum state $\rho_\lambda$ that is known as the probe state.
This state is then subjected to measurements and $\lambda$ is estimated from the outcomes with the aid of an estimator function.
The measurement can be described by a complete set of positive operator-valued measurement (POVM) operators $\{\Pi_n\}$ where the $n$-th outcome occurs with probability $p_n(\lambda) {=} \text{Tr}\left[\rho_\lambda \Pi_n\right]$.
This classical probability gives a statistical lower bound for the accuracy of estimation, given by the standard deviation $\delta \lambda$, in terms of the Cram\'er-Rao inequality $\delta \lambda \ge 1/\sqrt{\mathcal{M} F^C}$.
Here, the  number of measurement repetitions is $\mathcal{M}$ and the classical Fisher information (CFI), $F^C {=} \sum_n p_n (\partial_\lambda\ln p_n)^2$ depends on the measurement basis by construction~\cite{paris2009quantum}. 
The ultimate lower bound is given by the quantum Fisher information (QFI) $F^Q$, which is the maximum of the CFI over all possible choices of basis and thus, is a basis-independent quantity.
One can therefore define a quantum Cram\'er-Rao bound
\begin{align}\label{eq:cramer-rao}
\delta \lambda \ge \frac{1}{\sqrt{\mathcal{M} F^C}} \ge \frac{1}{\sqrt{\mathcal{M} F^Q}} \,.
\end{align}
To define QFI, one can consider the symmetric logarithmic derivative (SLD) operator $\mathcal{L}$, implicitly defined as $\partial_{\lambda}\rho_{\lambda} {=} (\rho_\lambda \mathcal{L}_\lambda {+} \mathcal{L}_\lambda \rho_\lambda)/2$. 
The QFI can be written as $F^Q {=} \text{Tr}\left[\mathcal{L}_{\lambda}^2 \rho_\lambda \right]$.
For pure states  $\rho_{\lambda} {=} \ket{\psi_{\lambda}} \bra{\psi_{\lambda}}$, one can arrive at a simplified version $\mathcal{L}_{\lambda} {=} 2 \partial_{\lambda} \rho_{\lambda}$, and therefore~\citep{paris2009quantum}
\begin{align}
F^Q = 4\left(\braket{\partial_\lambda \psi_{\lambda}|\partial_\lambda \psi_{\lambda}} - |\braket{\partial_\lambda \psi_{\lambda}|\psi_{\lambda}}|^2 \right) \,.
\label{QFI_eq}
\end{align}
QFI gives the ultimate precision limit on the estimation and the optimal basis to obtain this bound is not unique.
However, one choice of the optimal basis is always given by the projectors formed from the eigenvectors of the SLD operator.

Estimation theory can also be applied to NH systems. 
In such systems, the eigenstates do not form  an orthonormal basis set. 
This is due to the fact that the right and left eigenstates are not the same in the NH case as they are in the Hermitian scenario.
For a NH Hamiltonian $H_{\rm NH}$, they are defined as
\begin{align}
    H_{\rm NH} \ket{\psi^{\rm R}_n} &= E_n \ket{\psi^{\rm R}_n}, \nonumber \\
    \bra{\psi^{\rm L}_n} H_{\rm NH} &= E_n \bra{\psi^{\rm L}_n} \implies H_{\rm NH}^{\dag} \ket{\psi^{\rm L}_n} = E^*_n \ket{\psi^{\rm L}_n} ,
\end{align} 
with the corresponding eigenvalue $E_n$.
In conjunction with the left eigenvectors these states are biorthogonal~\cite{brody2014biorthogonal}, and upon normalization give, $\braket{\tilde{\psi}^{\rm L}_m | \tilde{\psi}^{\rm R}_n} {=} \delta_{mn}$.
Here, $\ket{\tilde{\psi}^{\rm R}_n} {=} \ket{\psi^{\rm R}_n} / \sqrt{\braket{\psi^{\rm L}_n | \psi^{\rm R}_n}}$, and $\ket{\tilde{\psi}^{\rm L}_n} {=} \ket{\psi^{\rm L}_n} / \sqrt{\braket{\psi^{\rm L}_n | \psi^{\rm R}_n}} \, ^*$.
To ensure that the measurement outcomes generate a normalized probability distribution, these states need to be divided by their norms calculated with standard inner products. 
This means, to describe the probe corresponding to the right-eigenstate with a valid density operator one has to consider $\ket{\tilde{\psi}^{\rm R}_n}\bra{\tilde{\psi}^{\rm R}_n} / \text{Tr}(\ket{\tilde{\psi}^{\rm R}_n}\bra{\tilde{\psi}^{\rm R}_n})$.
For a dynamical probe state generated by the non-Unitary operator $U {=} e^{-i _H{\rm NH} t / \hbar}$ on an initial state $\ket{\psi_0}$, the normalized density operator is $U \ket{\psi_0} \bra{\psi_0} U^{\dag} / \text{Tr}(U \ket{\psi_0} \bra{\psi_0} U^{\dag})$.
This is a standard approach for pure state probes in NH settings which has been explored  both theoretically~\cite{alipor2014quantum, yu2023quantum} and experimentally~\cite{xiao2024non, yu2024toward}.
This gives rise to a valid density operator with which the standard QFI definition~\cite{paris2009quantum, braunstein1994statistical} in Eq.~\eqref{QFI_eq} can be applied.

\section{Model} 
\label{Model}

We consider the 1D tight-binding model on a $L$-site system with a linear Stark field
\begin{align}
    H_{\rm S} = \sum_{j} J \left(\ket{j} \bra{j{+}1} + \ket{j{+}1} \bra{j}\right) + \sum_{j} h j \ket{j} \bra{j},
    \label{eq:Stark_ham}
\end{align}
where $J$ is the nearest-neighbor tunneling parameter and $h$ is the linear field.
In the absence of the field, the eigenstates of $H_{\rm S}$ are completely delocalized and are given by the Bloch wavefunctions.
As $h$ is turned on, above a critical value $h_c$, namely $h {\ge} h_c$, all the eigenstates are localized at different sites.
In the thermodynamic limit, the eigenfunctions can be expressed in terms of spherical Bessel functions and the eigenenergies are equally spaced with steps of $h$~\cite{hartmann2004dynamics}.
This is known as the Wannier-Stark localization~\cite{kolovsky2003bloch}.
In the thermodynamic limit the critical $h$ value is zero, but for finite systems the critical parameter is nonzero.
Below the critical point, the system is in the so-called extended phase.
The eigenstates in the extended phase are extremely sensitive to changes in $h$, and the QFI with respect to $h$ for all the eigenstates reveal quantum-enhanced sensitivity, i.e., $F^Q(h) {\sim} L^{\beta}$ with $\beta {>} 2$~\cite{he2023stark}.
This makes them excellent probes for weak fields. 

As the whole spectrum is affected by the Stark localization phase transition and subsequently shows favorable scaling of the QFI, it seems plausible that the QFI would display quantum advantage during non-equilibrium dynamics.
Indeed, in Ref.~\cite{manshouri2024quantum}, it was shown, by initializing a single particle in the middle of the lattice, that the long-time value attained by the QFI in the extended phase also reaches quantum-enhanced sensitivity along with the characteristic quadratic scaling with time~\cite{ilias2022criticality}.
In the single-excitation subspace, the critical value of $h$ separating the extended phase from the localized phase was shown to be $h_c{\approx} 8J/L$. 
Similarly, in the multi-excitation subspaces, the critical point $h_c$ decreases as the system size increases. 
This makes the ergodic phase vanishingly small for large system sizes and thus limits the quantum-enhanced sensitivity to finite probe sizes~\cite{manshouri2024quantum}. 

A potential physical platform for realizing the non-equilibrium Stark probe is cold atoms in optical lattices where the linear field term can be easily implemented~\cite{dahan1996bloch, anderson1998macroscopic, meinert2014interaction, preiss2015strongly, scherg2021observing, kohlert2023exploring}. 
Although these systems possess long coherence time, the presence of noise is unavoidable. 
This noise predominantly stems from spontaneous emissions, due to the atoms coupling to the vacuum modes of the electromagnetic field. 
A spontaneous emission effectively carries out a position measurement on the atom on a length scale given by the wave number of the emitted photon, which is typically in the optical range.
As a result, the atom is localized within a lattice site.
This causes dephasing by reducing the initially coherent atomic wavefunction to an incoherent mixture of localized Wannier states at different sites. 
The master equation for the atomic density operator can be derived based on the microscopic understanding for the mechanism of spontaneous emission under the suitable approximations, namely the dipole, Born-Markov, and rotating wave approximations.
For bosons loaded in the lowest Bloch band of the lattice, the resulting decoherence operators can be shown to be the number operators at each site~\cite{pichler2010nonequilibrium}. 
The lowest band approximation is based on the ultracold temperature and typical tight confinement of the lattice. 
The number operators are also relevant in experiments with two-species fermions, that are usually realized with two different hyperfine states of a particular atomic sample~\cite{gorshkov2010two}.
Due to the large detuning of the laser frequency (that generates the optical lattice) from the atomic transition frequency, the spontaneously emitted photons from these two species are practically identical. 
Therefore, the decohering mechanism at a particular site depends on the total number of particles at the site and not on the spin species~\cite{sarkar2014light}.
The provision of long coherence time in optical lattice setups also allows for engineered noise instead, such as photon emissions stimulated by a dedicated laser slightly detuned from relevant atomic transitions~\cite{luschen2017signatures}.
In this case, the decoherence strength can be controlled by changing the laser intensity.

In practice, independent of the physical platform, the presence of noise and deviation from perfect unitary evolution is inevitable. 
In this paper, we generalize the results of Ref.~\cite{manshouri2024quantum} to realistic scenarios and investigate the effect of noise  in estimating the gradient field $h$. 
As the result the evolution of the system will be given by open system quantum dynamics. 
We consider two different formulations of such dynamics, namely, (i) the Markovian master equation and (ii) Lindblad-like trace-preserving dynamics that is equivalent to evolution with a NH Hamiltonian. 
With these formulations, we investigate the impact of noise on sensing quality of the gradient field $h$.

\section{Markovian master equation dynamics}
\label{Dephase}

\begin{figure}[t]
\centering
\includegraphics[width=0.99\linewidth]{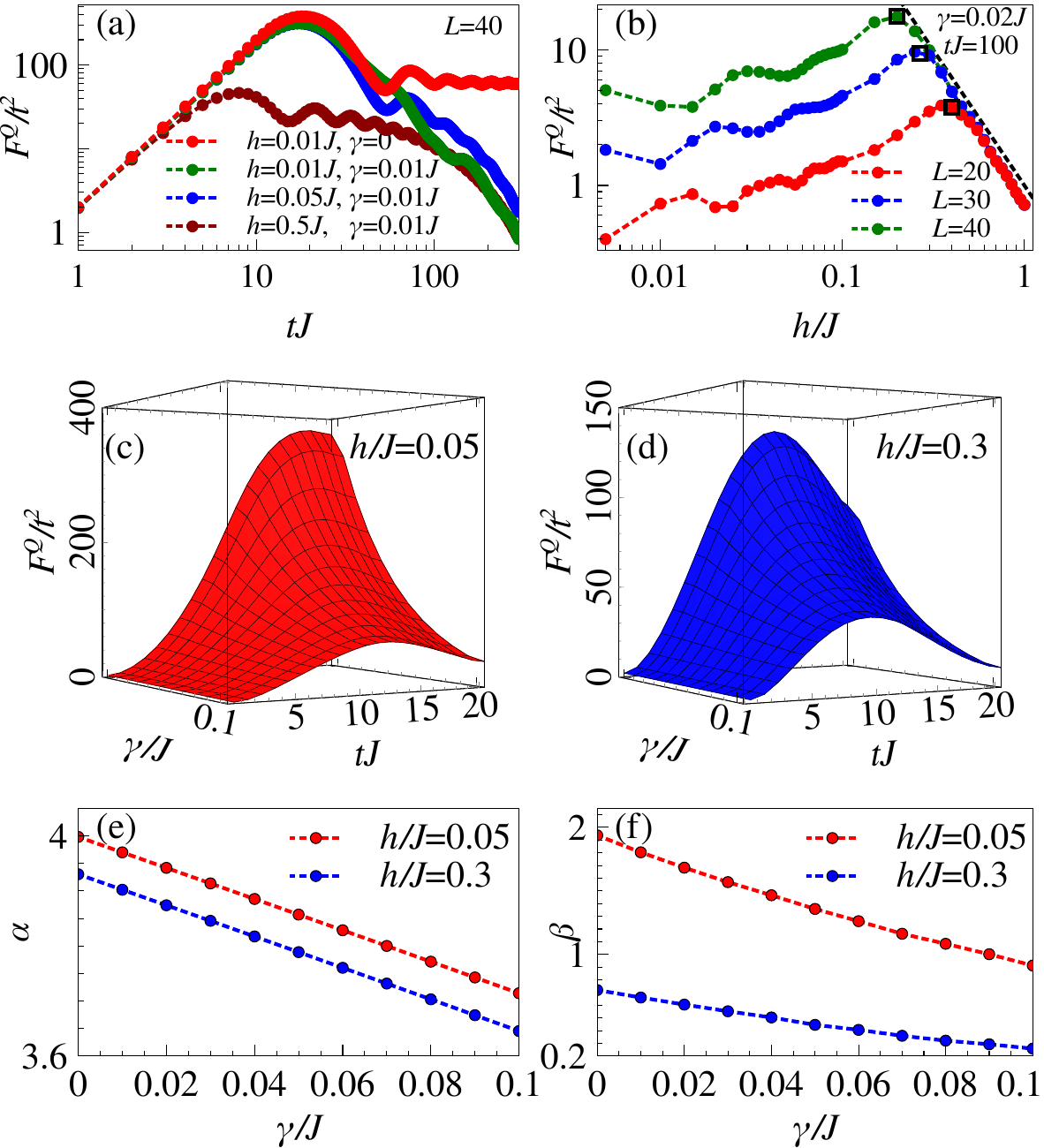} 
\caption{\textbf{Stark system dynamics with dephasing}. (a) Comparison of the evolution of $F^Q/t^2$ for $L {=} 40$ between the the decoherence-free case ($\gamma {=}$ 0) and $\gamma {=} 0.01 J$ for various field strengths $h$. (b) $F^Q/t^2$ at $tJ {=} 100$ for $\gamma {=} 0.02 J$ as a function of $h$ for various system sizes. Black squares show the transition points for different $L$. Dotted line is the $1/h^2$ fit. $F^Q/t^2$ as a function of decoherence strength $\gamma$ and time $t$ for $L {=} 40$ with (c) $h {=} 0.05J$ (extended phase) and (d) $h {=} 0.3J$ (localized phase). (e) Exponent $\alpha$ for scaling of QFI in short time, $F^Q {\sim} t^{\alpha}$. (f) Exponent for scaling with $L$ for the maximum value of $F^Q/t^2$.}
\label{Fig_Lind}
\end{figure}

Quantum superposition is very delicate and can quickly change into a classical mixture as the result of interaction with environment.
As mentioned above, the Lindblad operators can be the number operators at each site conveying the effect of decoherence. 
Therefore, the Lindblad form of the master equation for the system density operator $\rho$ can be written as
\begin{align}
\dot{\rho} = -\frac{i}{\hbar}[H_{\rm S},\rho] + \frac{\gamma}{2} \sum_{j} (2n_{j}\rho n_{j}^{\dag} - n_{j}^{\dag} n_{j}\rho - \rho n_{j}^{\dag} n_{j}),
\label{eq:master}
\end{align}
where $\gamma$ is the decoherence strength and the number operators $n_j$'s are the Lindblad operators. In particular, for the case of dephasing the Lindblad operators are given by $n_j{=} \ket{j} \bra{j}$.
One technique to compute the dynamics exactly is to coloumnwise vectorize the density matrix $\rho$ to a vector form $\tilde{\rho}$ and then rewrite the master equation as $\dot{\tilde{\rho}} = \tilde{\mathcal{L}} \tilde{\rho}$ with the Liouvillian operator accordingly modified as
\begin{align}
\tilde{\mathcal{L}} &= -\frac{i}{\hbar}[\mathds{1} \otimes H_{\rm S} - H_{\rm S}^T \otimes \mathds{1}] \nn &+ \frac{\gamma}{2} \sum_{j} (2 n_{j}^* \otimes n_{j} - \mathds{1} \otimes n_{j}^{\dag} n_{j} - (n _{j}^{\dag} n_{j})^T \otimes \mathds{1}) .
\label{eq:Louvillian}
\end{align}
Starting with the particle in the middle of the lattice, we now look at the evolution of QFI during the open system dynamics in the presence of dephasing.
In the presence of decoherence, the QFI first increases in time and then show a steady decrease.
This is in contrast to the quadratic scaling with time shown by the long-time QFI in absence of noise, which is characterized by a constant value attained by $F^Q/t^2$ in the long time.
This can be seen clearly in Fig.~\ref{Fig_Lind}(a) for a system $L {=} 40$.
In Fig.~\ref{Fig_Lind}(b), we show the distinct behaviors in the extended phase and the localized phase by plotting $F^Q/t^2$ at $tJ {=} 100$ for different system sizes for $\gamma {=} 0.02 J$.
We see that beyond $h {\approx} 8J/L$, the value of $F^Q/t^2$ becomes size independent and falls as ${\approx} 1/h^2$.
As the QFI decays to zero in the long time, we focus on analyzing the scaling behavior up to an intermediate time.
As Fig.~\ref{Fig_Lind}(c) shows, for different strengths of decoherence $\gamma$ the value of $F^Q/t^2$ reaches a maximum at a short time before it starts decreasing.
The system size is $L {=} 40$ here which means the system is in the extended phase in Fig.~\ref{Fig_Lind}(c).
In Fig.~\ref{Fig_Lind}(d) the localized phase is considered as $h$ is increased to $0.3J$.
This shows a similar behavior albeit reducing the QFI values overall.
We analyze the scaling of $F^Q$ in the short-time limit, i.e., $F^Q {\sim} t^{\alpha}$, where $\alpha {=} 4$ in absence of decoherence~\cite{manshouri2024quantum}.
As is displayed in Fig.~\ref{Fig_Lind}(e), although $\alpha$ decreases with increase in $\gamma$, the drop is not drastic.
This shows that the scaling advantage is retained even for moderate amount of dephasing during the dynamics.
Finally, we look at the scaling of the maximum value of $F^Q/t^2$ attained with system size, i.e., $\text{max}(F^Q/t^2) {\sim} L^{\beta}$ in Fig.~\ref{Fig_Lind}(f).
In the extended phase the exponent $\beta$ decreases monotonically with $\gamma$ from the Heisenberg limited value of 2 towards the standard limited value of 1.
In the localized phase, no quantum advantage is observed as $\beta$ is always below 1, which is the expected behavior.

\section{Trajectory-Based Non-Hermitian Description}
\label{Traj}

An instructive way to comprehend the connection between the Lindblad master equation dynamics of the density operator and evolution of the state under a NH Hamiltonian is given by the quantum trajectory method. 
This method is often employed when the system size is too large for exact calculations of the master equation in Eq.~(\ref{eq:master}) due to the exponential growth of the Hilbert-space dimension with $L$, and is also referred to as the Monte Carlo wave-function method~\cite{daley2014quantum}. 
In this method, the master equation in Eq.~\eqref{eq:master} is first rewritten as 
\begin{align}
\dot{\rho} = -\frac{i}{\hbar} \left( H_{\text{eff}} \rho - \rho H_{\text{eff}}^{\dag} \right) + \gamma\displaystyle\sum_j   n_j \rho n_j^{\dag} , 
\label{eq:q_traj} 
\end{align}
with an effective NH Hamiltonian defined as 
\begin{align}
H_{\text{eff}} = H_{\rm S} - \frac{i\gamma}{2} \sum_j n_j^{\dag} n_j . 
\label{eq:Heff}
\end{align}
The final state in this approach is a stochastic average over individual trajectories which are numerically evolved as pure states. 
Each trajectory is constructed by stochastic action of the Lindblad operators and evolution under the effective NH Hamiltonian $H_{\rm eff}$ in the following steps:
\begin{enumerate} [leftmargin=*]
    \item The pure state $\ket{\psi(t)}$ is evolved non-Unitarily for a short time $\Delta t$ to $e^{-iH_{\rm eff}\Delta t/\hbar} \ket{\psi(t)}$.
    
    \item The resulting loss in norm $\Delta p$ can be attributed to the total action of all possible Lindblad operators as $\Delta p = 1 - \braket{\psi(t) | e^{iH_{\rm eff}^{\dag}\Delta t/\hbar} e^{-iH_{\rm eff}\Delta t/\hbar} | \psi(t)} = \gamma \Delta t \sum_j \braket{\psi(t) | n_j^{\dag} n_j | \psi(t)} \coloneqq \Delta t \sum_j p_j$. The probability that a particular Lindblad operator $n_j$ acts is $p_j \Delta t/\Delta p$. The normalized state at time $t+\Delta t$ is chosen to be
    \begin{align}
    \ket{\psi(t+\Delta t)}= \left\{ \begin{array}{cc} 
    \frac{e^{-iH_{\rm eff}\Delta t/\hbar} \ket{\psi(t)}}{\sqrt{1-\Delta p}}  & \text{with probability } 1-\Delta p \\
    \frac{\sqrt{\gamma} n_j \ket{\psi(t)}}{\sqrt{\Delta p_j}} & \text{with probability } \Delta p 
    \end{array} \right. \nonumber
    \end{align}
    
    \item The stochastic choices of evolutions with $H_{\rm eff}$ and action of particular Lindblad operators at different time instances give rise to a trajectory of the state. Averaging over a large number of trajectories recovers the dynamics under the master equation in Eq.~\eqref{eq:master} up to first order in $\Delta t$. This can be seen by looking at the rate of change of the density operators corresponding to $\ket{\psi(t)}$ and $\ket{\psi(t+\Delta t)}$.    
\end{enumerate}

The importance of NH Hamiltonians in open system dynamics is thus showcased by this method.
A special trajectory is given by the scenario where no Lindblad operator $n_j$ acts on the system and therefore, is only evolved with $H_{\rm eff}$.  
In this situation, the dynamics is given by the pure state $\ket{\psi(t)} {=} e^{-iH_{\rm eff}t/\hbar} \ket{\psi(0)} / \sqrt{p(t)}$, with $\ket{\psi(0)}$ as the initial state of the system.
Here $p(t) {=} \braket{\psi(0) | e^{iH_{\rm eff}^{\dag}t/\hbar} e^{-iH_{\rm eff}t/\hbar} | \psi(0)}$ is the probability to observe such trajectory.
As this probability decays exponentially with time, the NH description is more appropriate in a short-time limit. 

In the case of the dephasing, namely $n_j{=}\ket{j}\bra{j}$, the effective Hamiltonian $H_{\rm eff}$, see Eq.~(\ref{eq:Heff}), only differs from $H_{\rm S}$ by a constant shift in the diagonal elements in the system.
Therefore, they have the same eigenvectors and eigenvalues would only differ by $i \gamma/2$.
This would result in identical dynamics, given that the state is normalized at each time step to obtain a valid probability distribution.
We will therefore exemplify the sensing of the Stark field in the NH description by exploiting two slightly different scenarios in the following section.

\section{Lindblad-like Trace-Preserving Non-Hermitian Description}
\label{NH}

The trajectory-based NH description, discussed in the previous section, is based on NH Hamiltonians the eigenvalues of which have negative imaginary components. 
For a more general realization of NH dynamics, one can engineer incoherent addition or subtraction of particles during the evolution under a number conserving Hermitian Hamiltonian. 
The resulting non-Unitary evolution does not conserve the number of particles and energy of the system. 
The NH Hamiltonians generating the evolution have eigenvalues tht are arbitrary complex numbers. 
In the presence of parity-time (PT) symmetry, the eigenvalues become real in PT-unbroken phase. 
In the PT-broken phase, the eigenvalues appear as complex conjugate pairs such as $E_n{=}E_n^{(R)} {\pm} i E_n^{(I)}$ (with $E_n^{(I)}\ge 0 $). 
During the evolution, the contribution of the eigenvectors with positive imaginary component in the eigenvalue is amplified (gain), while the contribution from the eigenvectors with negative imaginary component in the eigenvalue is diminished (loss). 
These types of evolution have been experimentally realized in various platforms such as electric circuits~\cite{Helbig2020Generalized, Hofmann2020Reciprocal, Liu2021Non}, acoustic~\cite{Zhang2021Acoustic, Gao2022Anomalous} and photonic lattices~\cite{Weidemann2020Topological, Song2020Two}, mechanical metamaterials~\cite{Brandenbourger2019Non, Ghatak2020Observation}, optical lattices~\cite{Lapp2019Engineering, Gou2020Tunable}, and photonic quantum walks~\cite{xiao2017observation, xiao2019observation, xiao2021observation}. 
The gain and loss terms can also be thought of as non-reciprocal couplings in the NH Hamiltonian which will be exemplified in this section. 

Note that unlike the trajectory-based dynamics which is valid only for short timescales, here the evolution is exact under the action of a NH Hamiltonian $H_{\rm NH}$. 
In this formalism the non-preserving norm of the dynamics can be interpreted as the exchange of energy or particles with environment. 
In fact, this dynamics can  be shown to be governed by a Lindblad-like master equation, as prescribed in Ref.~\cite{brody2012mixed}. 
By decomposing $H_{\rm NH} {=} H_{\rm h} {-} i \gamma H_{\rm ah}$ into a Hermitian part $H_{\rm h}$ and an anti-Hermitian part $-i \gamma H_{\rm ah}$ (here $H_{\rm ah}^{\dag} {=} H_{\rm ah}$) that is associated with a decoherence rate $\gamma$, the dynamical equation for $\rho$ can be written as 
\begin{align}
\dot{\rho} = -\frac{i}{\hbar} [H_{\rm h}, \rho] + \gamma \left(2\text{Tr}(\rho H_{\rm ah}) \rho - H_{\rm ah} \rho - \rho H_{\rm ah} \right) .
\label{eq:master_trace}
\end{align}
This type of evolution has been utilized before to study field-driven phase transitions in dissipative systems~\cite{tripathi2016parity} and current-driven steady state entanglement~\cite{panda2020entanglement}.
It is easy to see that under Eq.~\eqref{eq:master_trace}, the density operator preserves its positivity and norm (i.e., $d/dt \text{Tr}\rho {=} 0$).
The purity $d/dt \text{Tr}\rho^2$ is not conserved in general during the evolution as $d/dt \text{Tr}\rho^2 {=} -4 [\text{Tr}(H_{\rm ah} \rho^2) -\text{Tr}(H_{\rm ah} \rho) \text{Tr}(\rho^2)]$.
However, an initial pure state stays pure during the time evolution~\cite{brody2012mixed}.
The time evolution of the density operator can be written as $\rho(t) {=} e^{-iH_{\rm NH}t/\hbar} \rho_0 e^{iH_{\rm NH}^{\dag}t/\hbar}/\text{Tr}(^{-iH_{\rm NH}t/\hbar} \rho_0 e^{iH_{\rm NH}^{\dag}t/\hbar})$, where $\rho_0$ is the initial density operator.
In particular, for an initial pure state $\rho_0 {=} \ket{\psi_0} \bra{\psi_0}$, the evolved state remains pure and is dynamically given by 
\begin{align}\label{eq:time_evolution_NH}
\ket{\psi(t)} = \frac{\sum_n e^{-i E_n t/\hbar} \braket{\psi^{\rm L}_n | \psi_0} \ket{\psi^{\rm R}_n}}{\| \sum_n e^{-i E_n t/\hbar} \braket{\psi^{\rm L}_n | \psi_0} \ket{\psi^{\rm R}_n} \|} ,
\end{align}
where $\ket{\psi^{\rm R}_n}$ and $\bra{\psi^{\rm L}_n}$ are the right and the left eigenvectors of $H_{\rm NH} {=} H_{\rm h} - i \gamma H_{\rm ah}$, respectively,  with eigenvalue $E_n$ and $\|.\|$ denoting the vector norm.
This enables us to examine the dynamical properties of $H_{\rm NH}$ from a sensing perspective whereas the static features can be understood by analyzing the eigenstates. 

\subsection{Example 1: Nonreciprocal lattice}
\label{HN}

\begin{figure}[t]
\centering
\includegraphics[width=0.99\linewidth]{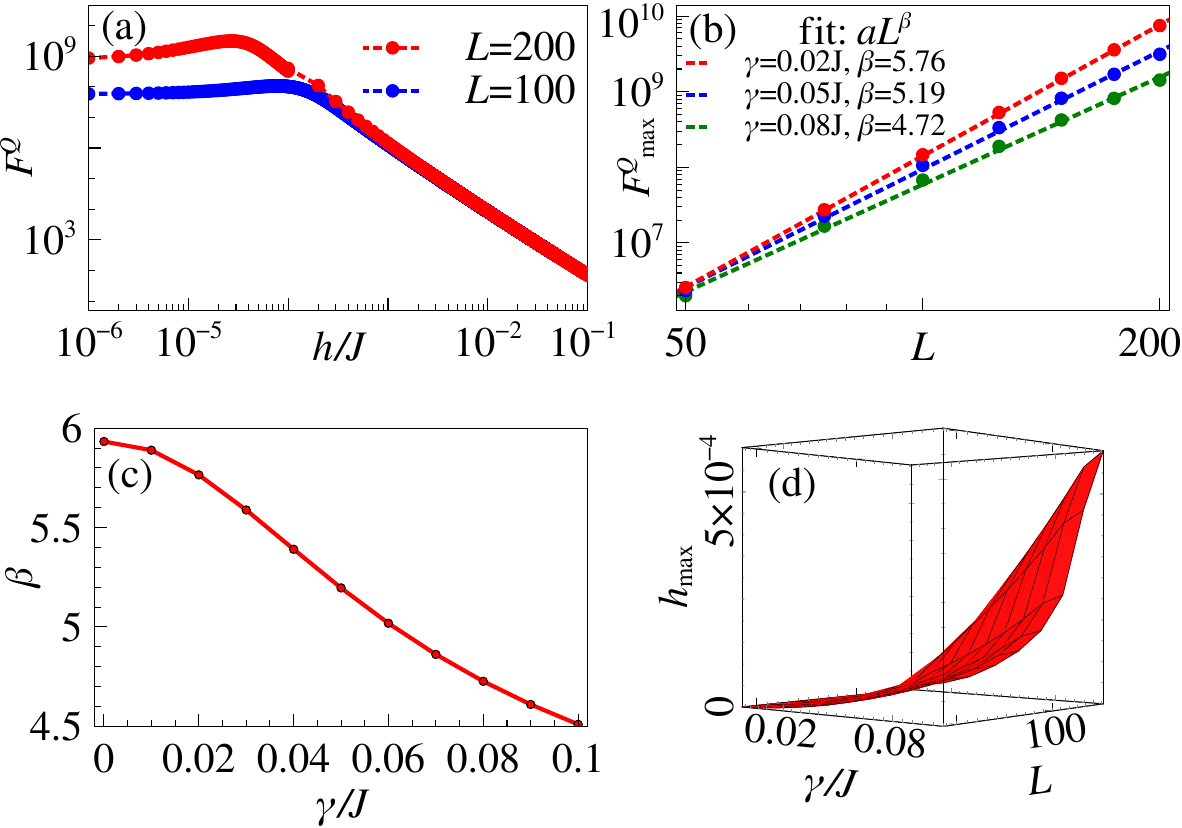} 
\caption{\textbf{Static sensing with NH Hatano-Nelson Hamiltonian}. (a) QFI of the lowest (real) energy state for $\gamma {=} 0.05$. (b) Maximum QFI attained for different $\gamma$ values as function of $L$.  (c) Scaling exponent of maximum QFI in the extended phase with $L$. (d) The value of $h$ where the QFI attains the maximum as a function of $\gamma$ and $L$.}
\label{Fig_HN1}
\end{figure}

In order to apply the formalism stated above to construct NH probes for sensing the Stark field, we adapt the driven spin system studied in Ref.~\cite{panda2020entanglement} to a spinless fermion on a lattice with gradient field.
Specifically, we consider $H_{\rm h} {=} \sum_{j} J \cosh{\mu} \, (\ket{j} \bra{j{+}1} + \ket{j{+}1} \bra{j}) + \sum_{j} h j \ket{j} \bra{j}$ with $\mu {=} \sinh^{-1}{\gamma}$, and  $H_{\rm ah} {=} i \sum_{j} J (\ket{j} \bra{j{+}1} {-} \ket{j{+}1} \bra{j}) $.
We can then describe the evolution exactly with an effective NH Hamiltonian $H_{\rm h} - i \gamma H_{\rm ah}$, which takes the form of the famous Hatano-Nelson Hamiltonian with nonreciprocal tunneling~\cite{Hatano1996Localization, Hatano1997Vortex}
\begin{align}
    H_{\rm HN} = \sum_{j} \left( J_L \ket{j} \bra{j{+}1} + J_R \ket{j{+}1} \bra{j}\right)  + \sum_{j} h j \ket{j} \bra{j} ,
    \label{eq:HN_ham}
\end{align}
where the tunneling parameters towards left and right are $J_L {=} Je^{\mu}$ and $J_R {=} Je^{-\mu}$, respectively.
This model has been experimentally realized on several platforms, among which the superconducting qubit based dynamical implementation~\cite{shen2025observation} is quite suitable for the task at hand as the gradient field can be applied trivially.
Here the non-Unitarity and nonreciprocity stem from strategic post-selection of multiple ancilla qubits.
Other realizations include a photonic platform with multiple frequency modes on a ring resonator serving as the sites with detuned coupling between the modes giving rise to the nonreciprocity~\cite{wang2021generating}, and acoustic phononic crystals with passive losses~\cite{gu2022transient}.
Schemes based on optical lattices with engineered dissipation have also been proposed~\cite{gong2018topological}.

For a finite system, the eigenvalues of $H_{\rm HN}$ are real and thus can be ordered.
When $J_L {\neq} J_R$ i.e.~$\gamma {\neq} 0$ and the gradient field term is absent, all the eigenstates are localized on the left edge of the system for $\gamma {>} 0$.
This is known as the NH skin effect~\cite{yao2018edge, martinez2018non}.
When the field term is turned on, it tries to localize different eigenstates at different sites.
The competition between the two localizing mechanisms has the strongest effect on the extremum eigenstates which become extremely sensitive to minute changes in $h$.
Indeed, we observe that the QFI with respect to $h$ in the ground state reaches high values, as shown in Fig.~\ref{Fig_HN1}(a).
We also note that the QFI reaches a maximum right before the field term overwhelms the skin effect.
This maximum $F^Q_{\rm max}$ is shown in Fig.~\ref{Fig_HN1}(b) as $L$ is varied for different $\gamma$ values.
For a given $\gamma$, this maximum shows an algebraic scaling with $F^Q_{\rm max} {\sim} L^{\beta}$. 
In Fig.~\ref{Fig_HN1}(c), the exponent $\beta$ is shown as a function of the decoherence strength $\gamma$ which clearly shows strong quantum-enhanced sensing. 
The value of $h$ where the QFI attains its maximum is shown in Fig.~\ref{Fig_HN1}(d) as a function of $\gamma$ and $L$.
This $h_{\rm max}$ value is observed to be monotonically increasing with increasing $\gamma$.
This suggests that the range of $h$ over which high QFI values are attained can be broadened with decoherence strength.

\begin{figure}[t]
\centering
\includegraphics[width=0.99\linewidth]{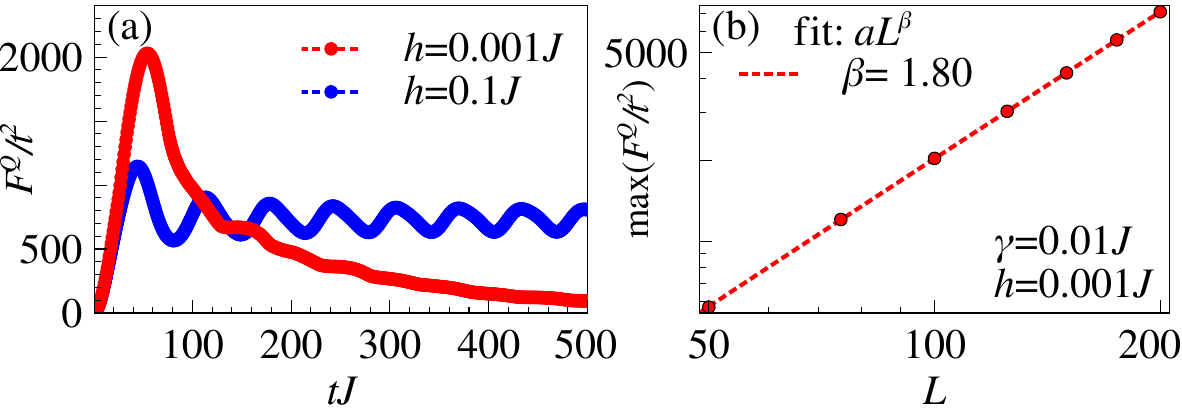} 
\caption{\textbf{Dynamic sensing with NH Hatano-Nelson Hamiltonian}. (a) Time evolution of $F^Q/t^2$ in the extended phase ($h {=} 0.001J$) and localized phase ($h {=} 0.1J$) for $L {=} 100$ and $\gamma {=} 0.05$. (b) Exponent for scaling with $L$ for the maximum value of $F^Q/t^2$ in the extended phase.}
\label{Fig_HN2}
\end{figure}

Beyond the ground state of the Hamiltonian~\eqref{eq:HN_ham}, one can also use the probe under non-equilibrium dynamics. 
In this situation, the probe is prepared by a single particle initially localized at the middle of the lattice. The evolution of the system is then given by Eq.~\eqref{eq:time_evolution_NH}. 
One can use the quantum state $\ket{\psi(t)}$ for estimating the value of $h$.  
Time evolution of $F^Q/t^2$ is shown in Fig.~\ref{Fig_HN2}(a) for $\gamma {=} 0.05J$ and a $100$ site lattice.
For the weak field ($h {=} 0.001J$), the value of $F^Q/t^2$ reaches a maximum in short time which shows super-linear scaling with system size (Fig.~\ref{Fig_HN2}(b)). 
In the field-localized phase ($h {=} 0.1J$) we observe that $F^Q/t^2$ moves towards a scale-invariant constant value in the long-time limit in an oscillating manner.

\subsection{Example 2: Unidirectional lattice}
\label{Lattice}

\begin{figure}[t]
\centering
\includegraphics[width=0.99\linewidth]{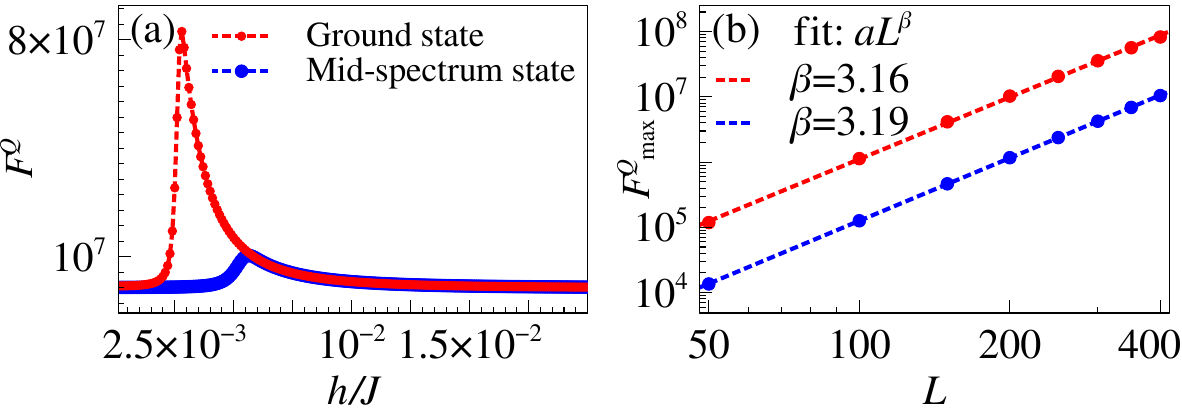} 
\caption{\textbf{Static sensing with NH unidirectional lattice Hamiltonian}. (a) QFI of the ground state and the mid-spectrum state for $L {=} 400$. (b) Exponent for scaling with $L$ for the maximum $F^Q$ value. }
\label{Fig_Uni1}
\end{figure}

After considering the previous NH description of an exact open system dynamics, we now look at evolution under another NH Hamiltonian as an effective description of noisy evolution.
The case that we consider is a unidirectional lattice, given by 
\begin{align}
    H_{\rm uni} = \sum_{j} J \ket{j} \bra{j{+}1} + \sum_{j} h j \ket{j} \bra{j} \,.
    \label{Uni_ham}
\end{align}
Here, only hopping to the left with strength $J$ is allowed.
Although this system can be thought of a special case of the previous example, the complete suppression of hopping in one direction gives rise to interesting sensing properties as we highlight in the following.
Unidirectionality has been experimentally realized in different platforms such as photonic lattices~\cite{regensburger2012parity} and coupled optical resonators~\cite{peng2014parity, peng2016chiral}. 
There are also proposals based on time reversal symmetry breaking in 1D photonic lattices~\cite{longhi2015robust} and imaginary gauge field in 1D tight-binding lattices~\cite{longhi2015non}.
In the context of sensing a gradient field, two possible physical realizations of NH lattices with unidirectional hopping have been suggested in~\cite{longhi2014exceptional}.
The first one is based on light transport in engineered optical waveguide lattices with segmented regions of alternating gain and loss along with high and low index contrast regions.
This corresponds to a Hermitian hopping model with site-dependent complex potential that is periodically modulated in time.
With high enough modulation frequency, the rotating wave approximation gives the desired unidirectional Hamiltonian for judicious choice of the potential.
The gradient field in the transverse waveguide direction is implemented by circularly curving along axial propagation.
The second approach for implementation of the  NH model uses light dynamics in a mode-locked laser with sinusoidally driven intracavity amplitude and phase modulators.
Choosing both modulation depths to be equal implies unidirectionality in the dynamical equation of the amplitudes of the cavity modes for specific values of phase difference between the modulators.
The detuning between the modulation frequency and the cavity mode sets the gradient field strength.
Now for the Hamiltonian in Eq.~\eqref{Uni_ham}, the eigenenergies are real and take evenly spaced values in the steps of $h$ i.e.~$E_n {=} nh$.
The corresponding right eigenvectors, $\ket{\psi_n^R} {=} \sum_j c_{n,j} \ket{j}$, are given by the coefficients~\cite{longhi2014exceptional}
\begin{equation}
c_{n,j} = \left\{
\begin{array}{cc}
0 & j \geq n+1 \\
1 & j=n \\
\left( \frac{J}{h} \right)^{n-j} \frac{1}{(n-j)!} & 0 \leq  j < n
\end{array}
\right.
\end{equation}
The real and equally spaced spectrum is similar to the Hermitian case and we observe that the QFI of all the eigenstates with respect to $h$ show quantum-enhanced sensitivity.
As shown in Fig.~\ref{Fig_Uni1}(a) for the lowest energy and mid-spectrum energy state, the QFI peaks at a certain weak field value in the extended phase and they collapse onto each other for strong fields in the localized phase.
By plotting the maximum QFI, represented by $F^Q_{\rm max}$, as a function of system size $L$ in Fig.~\ref{Fig_Uni1}(b), one finds algebraic scaling as $F^Q_{\rm max} {\sim} L^\beta$, with $\beta {>} 3$ showing quantum-enhanced scaling. 

\begin{figure}[t]
\centering
\includegraphics[width=0.99\linewidth]{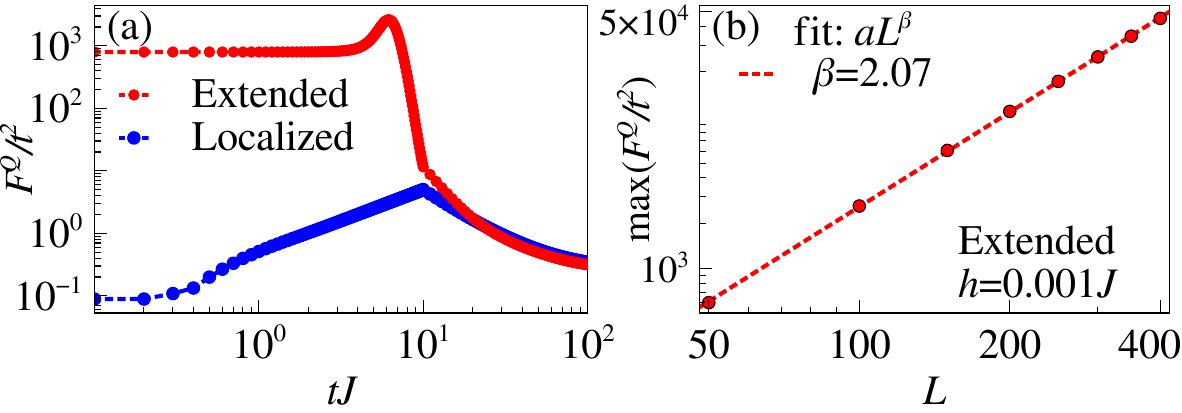} 
\caption{\textbf{Dynamic sensing with NH unidirectional lattice Hamiltonian}. (a) Time evolution of $F^Q/t^2$ in the extended phase ($h {=} 0.001J$) and localized phase ($h {=} 0.1J$) for $L {=} 100$. (b) Exponent for scaling with length for the maximum value of $F^Q/t^2$ in the extended phase.}
\label{Fig_Uni2}
\end{figure}

As all the eigenstates show such scaling with system size, one may wonder what happens if we use a non-equilibrium state of the system for inferring the unknown parameter $h$.  
We then look at the QFI dynamically by starting with the particle as a wave packet in the middle of the lattice i.e., $\ket{\psi(0)} = \sum_j \mathcal{N} e^{-(j-L/2)^2/2\sigma^2} \ket{j}$, with the normalization factor $\mathcal{N}$.
Then the evolution of the system is given by the non-Unitary operator $e^{-i H_{\rm uni} t/\hbar}$.
In this system, a periodic dynamics like Bloch oscillation is observed even if the lattice is finite and $h$ is weak~\cite{longhi2014exceptional}.
This is markedly different from the Hermitian case, where the periodic motion is halted when the wave packet reaches the lattice boundary and starts smearing out.
In the unidirectional lattice however, the periodic motion continues with time period $T {=} 2\pi/h$.
Time evolution of $F^Q/t^2$ in the extended phase ($h {=} 0.001J$) stays constant in the short-time limit, then reaches a maximum before decreasing steadily, as shown in Fig.~\ref{Fig_Uni2}(a).
For the localized phase case ($h {=} 0.1J$), we observe that $F^Q/t^2$ first increases before collapsing on the decreasing profile in the extended case at longer time. 
The exponent for scaling with $L$ for the maximum value of $F^Q/t^2$ in the extended phase is displayed in Fig.~\ref{Fig_Uni2}(b) and shows that the Heisenberg-limited precision can be achieved in this setup.

\section{Signal-to-noise ratio} 
\label{SNR}

As the Stark sensors are well suited for precision sensing of low gradient fields, one should also keep in mind the precision achieved during noisy dynamics with respect to the field strength. 
This can be quantified by the signal-to-noise ratio (SNR), defined as $h/\delta h$. 
By considering the optimal measurement and estimation algorithm, the Cram\'er-Rao bound in Eq.~\eqref{eq:cramer-rao} guarantees that $\delta h {\approx} 1/\sqrt{\mathcal{M} F^Q}$, which then implies that $\mathrm{SNR}\sim h\sqrt{\mathcal{M}F^Q}$. This clearly shows the difficulty of sensing small fields for which the SNR can only be enhanced by either increasing the sample size $\mathcal{M}$ or the quantum Fisher information $F^Q$. 
To see the effectiveness of Stark probes, we provide the SNR in Table~\ref{tab:precision} for the different case studies considered in this paper in both the extended and localized phases.
The SNR values are calculated for a typical sample size of $\mathcal{M}{=}1000$ and two different times, namely the optimal time $t_{\rm opt}$ at which $F^Q/t^2$ peaks or a fixed given time $t{=}10/J$. 
Note that, a SNR value greater than $10$ is experimentally acceptable for reliable estimation. 
The results, shown in Table~\ref{tab:precision}, are quite satisfactory in most cases, except for deep in the extended phase in the unidirectional lattice. 
In this case, more samples are needed to improve the SNR for a small field of $h/J{=}0.001$.
In addition, due to the favorable scaling with respect to system size, the precisions can be further improved by increasing the lattice size.
Note that, the system sizes considered in this paper are well within what can be achieved in experiments, for example, with tilted optical lattices~\cite{scherg2021observing, kohlert2023exploring}.

\begin{table}[t]   
\centering
\begin{tabular}{| c | c | c | c |}

\hline
Dynamical case & Phase & SNR ($t_{\rm opt}$) & SNR ($tJ{=}10$) \\
\hline

\multirow{3}{*}{Lindbladian} &  Extended ($h {=} 0.01J$) & $101.1$ & $42.2$ \\
\cline{2-4} 
&  Extended ($h {=} 0.05J$) & $474.4$ & $209.8$ \\
\cline{2-4} 
& Localized ($h {=} 0.5J$) & $861.7$ & $1016.7$ \\
\hline

\multirow{3}{*}{Nonreciprocal lattice} & Extended ($h{=}0.001J$) & $76.9$ & $4.3$ \\
\cline{2-4} 
& Extended ($h {=} 0.01J$) & $768.1$ & $43.4$ \\
\cline{2-4}
& Localized ($h {=} 0.1J$) & $4815.9$ & $423.5$ \\
\hline

\multirow{3}{*}{Unidirectional lattice} & Extended ($h {=} 0.001J$) & $10$ & $1.1$ \\
\cline{2-4} 
& Extended ($h {=} 0.01J$) & $96.9$ & $8.6$ \\
\cline{2-4}
& Localized ($h {=} 0.1J$) & $70.7$ & $70.7$ \\
\hline

\end{tabular}
\caption{SNR for the different examples considered in different phases with $\mathcal{M} {=} 1000$ measurement repetitions and $0.01J$ decoherence strength. Here, $L {=} 40$ for the Lindbladian evolution and $L {=} 100$ for the other two cases.}
\label{tab:precision}
\end{table}

\section{Conclusion}
\label{Conclusion}

Stark probes on a tight-binding lattice model with a gradient field are capable of achieving quantum-enhanced sensitivity with favorable scaling in terms of time and  system size during non-equilibrium dynamics with a starting state that can be easily initiated.
In the extended phase for weak fields, the QFI scales quartically in time at short times and quadratically in the long-time limit.
Recognizing their immense potential as quantum-enhanced sensors, we consider the practical implementation in the presence of decoherence that is unavoidable in practice.
We start with analyzing dephasing as a source of decoherence on the performance of sensing. 
While the quantum advantage of quartic scaling is only slightly degraded in short times, the quadratic scaling of long-time dynamics is lost. 
This sets an optimal time for sensing to be performed for obtaining the enhanced precision. 
In addition, the scaling with system size can stay above the standard limit in the extended phase for moderate decoherence strength.
This is quite an encouraging result for a non-equilibrium dynamics based sensor especially because the interferometric sensing protocols are acutely fragile to decoherence.
Along with the favorable scaling, we also observe satisfactory precision for the moderate system sizes considered in this paper.
We then establish the connection of such open system dynamics with evolution under a NH Hamiltonian.
While the quantum trajectory-based NH description approximates the true Lindblad master equation dynamics only at short timescales, we also provide a more general NH description which is not derived from a Lindblad master equation and describes an exact non-Unitary evolution engineered by incoherent gain and loss terms. 
Such systems can be described by an effective Lindblad-like master equation which interestingly prevents mixing the state of the system if it is initially pure. 
We provide two such examples to showcase the quantum-enhancement of sensing capability both for the eigenstates and dynamical states even in the presence of decoherence up to a certain strength.
Our paper demonstrates the applicability of Stark probes in the presence of quantum noise and discloses its close connection with NH sensors that can be realized in ongoing experiments.

\begin{acknowledgments}

SS acknowledges support from National Natural Science Foundation of China (Grant No.~W2433012).
AB acknowledges support from National Natural Science Foundation of China (Grants Nos.~12050410253, 92065115, and 12274059) and the Ministry of Science and Technology of China (Grant No.~QNJ2021167001L).

\end{acknowledgments}

\bibliography{Ref}

\end{document}